\documentclass[aps,prl,twocolumn,superscriptaddress]{revtex4-1}

\usepackage{graphicx}
\usepackage{epsfig}
\usepackage{amsmath,amsthm}
\usepackage{color}
\usepackage{verbatim}
\usepackage{ulem}

\begin{document}

\title{Magnetoresistance effects in the metallic antiferromagnet Mn$_2$Au}

\author{S.Yu.\,Bodnar}
\affiliation{Institut f\"{u}r Physik, Johannes Gutenberg-Universit\"{a}t, Staudingerweg 7, D-55099 Mainz, Germany}
\author{Y.\,Skourski}
\affiliation{Hochfeld-Magnetlabor Dresden (HLD-EMFL), Helmholtz-Zentrum Dresden-Rossendorf, 01328 Dresden, Germany}
\author{O.\,Gomonay}
\affiliation{Institut f\"{u}r Physik, Johannes Gutenberg-Universit\"{a}t, Staudingerweg 7, D-55099 Mainz, Germany}
\author{J.\,Sinova}
\affiliation{Institut f\"{u}r Physik, Johannes Gutenberg-Universit\"{a}t, Staudingerweg 7, D-55099 Mainz, Germany}
\author{M.\,Kl{\"a}ui}
\affiliation{Institut f\"{u}r Physik, Johannes Gutenberg-Universit\"{a}t, Staudingerweg 7, D-55099 Mainz, Germany}
\author{M.\,Jourdan}
\affiliation{Institut f\"{u}r Physik, Johannes Gutenberg-Universit\"{a}t, Staudingerweg 7, D-55099 Mainz, Germany}
\email{jourdan@uni-mainz.de}

\begin{abstract}
In antiferromagnetic spintronics, it is essential to separate the resistance modifications of purely magnetic origin from other effects generated by current pulses intended to switch the N{\'e}el vector. We investigate the magnetoresistance effects resulting from magnetic field induced reorientations of the staggered magnetization of epitaxial antiferromagnetic Mn$_2$Au(001) thin films. The samples were exposed to 60~T magnetic field pulses along different crystallographic in-plane directions of Mn$_2$Au(001), while their resistance was measured. For the staggered magnetization aligned via a spin-flop transition parallel to the easy [110]-direction, an ansiotropic magnetoresistance of $\simeq -0.15$~\% was measured. In the case of a forced alignment of the staggered magnetization parallel to the hard [100]-direction, evidence for a larger anisotropic magnetoresistance effect was found. Furthermore, transient resistance reductions of $\simeq 1$~\% were observed, which we associate with the annihilation of antiferromagnetic domain walls by the magnetic field pulses.

\end{abstract}



\maketitle

\section{I. INTRODUCTION}
In the emerging field of antiferromagnetic (AFM) spintronics~\cite{Jun18, Bal18, Jung18},  the orientation of the staggered magnetization, or more general of the N{\'e}el vector, is used to encode information. For devices, the manipulation of the N{\'e}el vector by current induced N{\'e}el spin-orbit torques (NSOTs) \cite{Zel14} is most promising and in principle the anisotropic magnetoresistance effect (AMR) can provide a read-out mechanism. Thus, current pulse induced reversible resistance changes demonstrated for CuMnAs \cite{Wad16, God18, Mat19} and Mn$_2$Au \cite{Bod18, Zho18, Mei18}, were interpreted as originating from the AMR associated with the reorientation of the N{\'e}el vector.

Current pulse generated  resistance switching was observed also in polycrystalline metallic AFM MnN/Pt bilayers \cite{Dun19} and insulating AFM NiO/Pt bilayers \cite{Che18,Mor18,Balt18}, which is assumed to result from interfacial spin-orbit torques and spin-Hall magnetoresistance for read-out.

However, recently similar current pulse induced resistance changes were obtained investigating non-magnetic cross structures consisting of pure Pt or Nb thin films. These experiments indicate that current pulse induced annealing could produce electrical signals similar to those observed in NiO/Pt \cite{Chi19, Mat19a}. Thus it is of major importance to investigate the obtainable magnetoresistance effects independent of current pulse based manipulation of the AFMs, i.\,e.\,to investigate resistance modifications purely associated with specific configurations of the N{\'e}el vector of antiferromagnetic domain configurations.

In this framework, we use a manipulation method of the N{\'e}el vector of Mn$_2$Au thin films, which guarantees that only magnetism related properties of the AFM are affected, explicitly excluding all kinds of Joule heating related effects: N{\'e}el vector reorientation by a large magnetic field pulse of up to 60~T. For a collinear AFM, it is energetically favorable, if the magnetic moments are aligned perpendicular to the magnetic field, which allows them to slightly cant towards the field direction. This allows us to study the magnetoresistance effects originating from a reorientation of the N{\'e}el vector excluding all possible spurious effects. 
In general one can consider three different magnetoresistance effects: First, the ordinary magnetoresistance (OMR), which appears in nonmagnetic metals as well. The field dependence of OMR is non-hysteretic, depends on the details of the band structure, and is in general positive \cite{Cal91}. The longitudinal as well as the transversal OMR can amount to several percent for large fields ($> 10$~T)  \cite{Lut59,Pip89}. Second, the anisotropic magnetoresistance (AMR), which, in AFM spintronics, is generally assumed to dominate the resistance modifications resulting from a current induced reorientation of the N{\'e}el vector. Third, the domain wall magnetoresistance (DWMR). This is a relevant effect in AFM spintronics, as typically the thin films are patterned into structures with lateral dimensions of several micrometers, whereas the typical AFM domain size is much smaller \cite{Sap18, Grz17, Wad18, Bod19}. In ferromagnets, the transport of spin-polarized currents leads to scattering by the non-collinear spin structure imparted by domain walls~\cite{Mar07, Bie13}. Similar effects are possible for AFMs as well \cite{Zhe20}, but have been studied much less. One of the few AFM metals, whose domain wall resistance has been experimentally investigated, is Cr \cite{Jar07}.

Here, we investigate the resistance changes induced by an reorientation of the N{\'e}el vector in epitaxial Mn$_2$Au(001) thin films by a magnetic field pulse, i.\,e.\,we investigate pure magnetoresistance effects excluding heating effects. We focus on the hysteretic magnetoresistance associated with the DWMR and AMR, as we are interested in the switching effects of AFM spintronics.

\section{II. EXPERIMENTAL TECHNIQUES}
For the electric transport measurements, an epitaxial Mn$_2$Au(001) thin film with a thickness of 80~nm was deposited on a Al$_2$O$_3$ (1$\bar{1}$02) substrate with a Ta(001) buffer layer (thickness 20~nm) as described in Ref.\,\cite{Jou15}. This sample was capped with 2~nm of Ta to prevent oxidation of the Mn$_2$Au surface and shows the same morphology as those investigated in Ref.\,\cite{Sap18}. For precise measurements of the resistance, it was patterned by optical lithography and ion beam etching into 3 stripes of 7~mm length and 200~${\rm \mu}$m width aligned parallel to the easy [110]-, easy [1$\bar{1}$0]- and hard [100]-directions of Mn$_2$Au, respectively. Separate contact pads at the ends of each stripe allowed for 4-probe measurements of the resistivity along the different crystallographic directions (see inset of Fig.\,1).

The sample was exposed to a magnetic field pulses with an amplitude of 60~T and a pulse duration of 150~ms at the high field laboratory of the Helmholtz center Dresden Rossendorf (HLD-EMFL) to generate an alignment of the N{\'e}el vector perpendicular to the field direction. To ensure the stable temperature required for magnetoresistance measurements, the sample was  immersed in liquid helium inside the cryostat within the field coil.  During each pulse, and afterwards for up to 10~s, the resistance $R$ of one of the patterned Mn$_2$Au stripes was probed with a sampling rate of 200~kHz using a numerical lock-in technique with a probe current of $10^5$~${\rm A/cm^2}$ modulated with a frequency of 20~kHz. In parallel, the magnetic field was obtained by numerical integration of the $dB/dt$ signal induced in a pick-up coil situated next to the sample. This lock-in technique enables resistance measurements during the field pulse application, while standard dc measurements are impossible due to the large induction effect generated by the magnetic field pulse. After each pulse, a waiting time of approximately 4~h is required for thermalization of the magnet coil, before the next pulse can be applied.

\section{III. RESULTS}
The magnetoresistance effects discussed below can only be understood, if we first summarize our previous microscopic investigations of the persistent effects of large magnetic field pulses on the AFM domain pattern of our samples \cite{Sap18}. We showed by X-ray magnetic linear dichroism - photoelectron emission microscopy (XMLD-PEEM) that the AFM domain pattern of as-grown Mn$_2$Au(001) thin films consists of domains with an area of $\simeq 1$~$\mu$m$^2$ with the N{\'e}el vector aligned along both easy [110]- and [1$\bar{1}$0]-axes. Exposing the samples to a field pulse with an amplitude of $B_{pulse}=30$~T along one easy axis resulted in a significantly increased area of AFM domains with the N{\'e}el vector aligned perpendicular to the field pulse direction. The sample area covered by domains with this N{\'e}el vector alignment was observed to saturate for $B_{pulse}=50$~T, i.\,e.\,an orientation of the N{\'e}el vector via spin-flop happens between 30~T and 50~T. We showed that these domains with the N{\'e}el vector aligned perpendicular to the field pulse direction are typically of sightly larger sizes than in the as-grown state of the samples and are separated by narrow wormlike spin structures with a width of $\simeq 100$~nm (about twice the resolution limit of the PEEM). However, after the application of field pulses along the hard [100]-direction, we observed an equal part of AFM domains with N{\'e}el vector aligned parallel to both easy [110]- and [1$\bar{1}$0]-axes, with as well a sightly larger sizes than in the as-grown state \cite{Sap18}.

From the resistance measurements presented here, we first discuss the effect of field pulses $\vec{B}(t)$ along an easy [110]-direction of Mn$_2$Au, as this is associated with a straight forward modification of the magnetic state. The pulse generates a spin flop of all AFM domains with the N{\'e}el vectors ${\vec N}\parallel\vec{B}$, i.\,e.\,results in an persistent alignment of ${\vec N}\perp{\vec B}$ as discussed above. This alignment results in a relatively small, but persistent AMR effect, as we will discuss below. However, we first present the larger resistance modifications associated with the magnetic field pulse. 
 
The red curve in Fig.\,1 shows the resistance $R$ of a Mn$_2$Au(001) thin film (normalized to the resistance $R_{0}$ before the field pulse) probed along the easy [110]-direction during a 150~ms pulse generating a time dependent magnetic field $B$(t) applied along the same direction.
\begin{figure}
\includegraphics[width=\columnwidth]{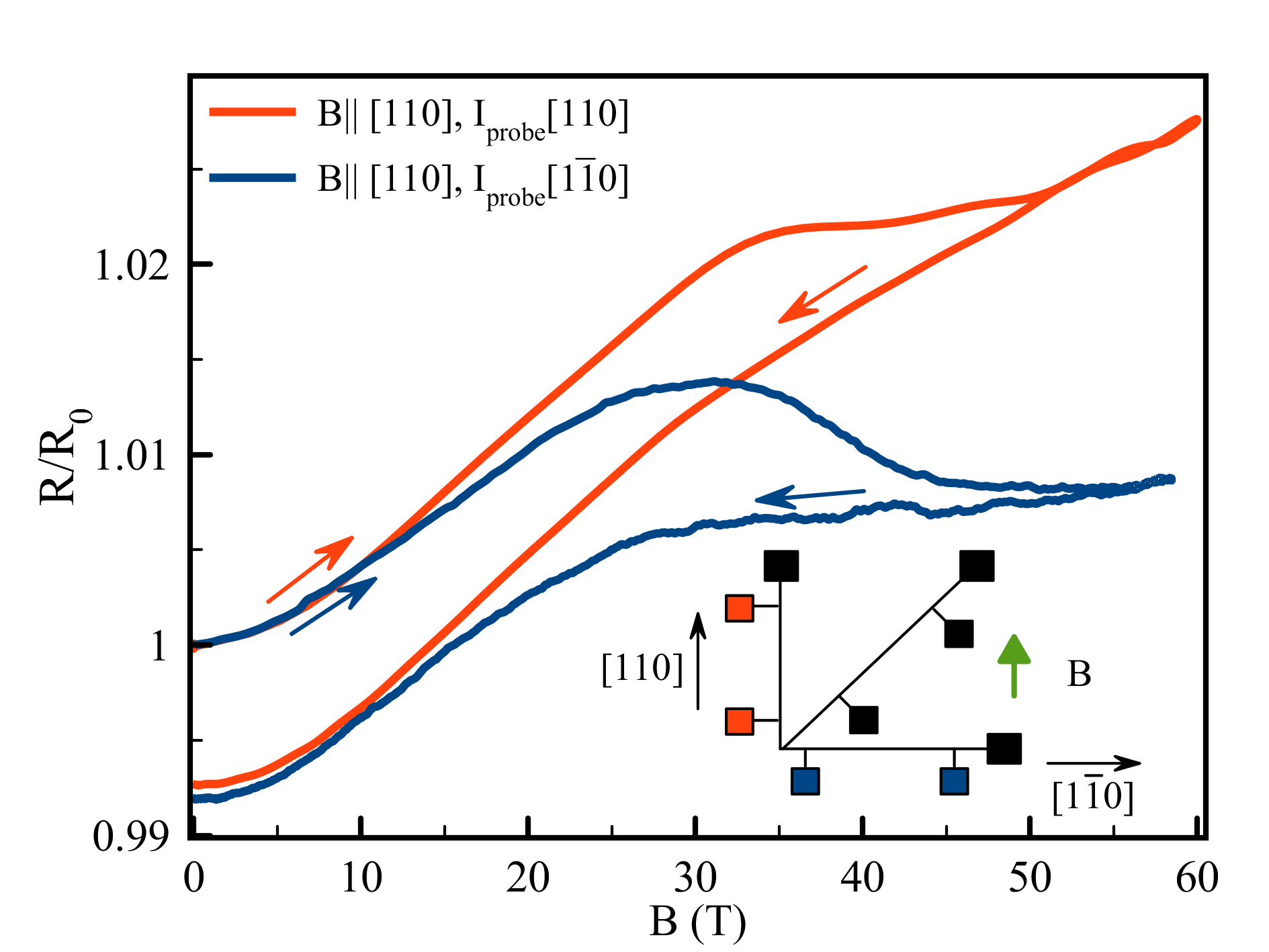}
\caption{\label{Fig2} 
Field dependent resistances of a Mn$_2$Au(001) epitaxial thin film measured during the exposure to magnetic field pulses along [110] with a duration of 150~ms, normalized to the sample resistances directly before the application of the respective field pulse. Between both field pulses a waiting time of about 4 h was required.}
\end{figure}

Up to $B(t)\simeq 30$~T, $R(B)$ is governed by the non-hysteretic positive OMR. However, we focus on the hysteretic resistance modifications originating from DWMR and AMR.  Such effects appear above 30~T, which is exactly the magnetic field at which we previously observed a spin-flop of most AFM domains with ${\vec N} \parallel {\vec B} \parallel $[110]. The hysteretic resistance reduction saturates at $\simeq$50~T, which is the magnetic field, at which the N{\'e}el vectors of all corresponding domains have completed the spin-flop alignment \cite{Sap18}. With decreasing field below 30~T, $R(B)$ hysteretically reproduces the previous curve with a negative shift of the resistivity of $\simeq 0.75$~\%, by which the sample resistivity directly after the pulse is reduced compared to its initial value. 

As we will discuss below, this resistance reduction relaxes on a time scale of $\simeq 10$~s. So if the experiment is repeated some hours later with another field pulse, one starts with the same resistance, which the as-grown sample had before the first field pulse plus a small AMR contribution, which we will discuss below. However, first we continue to discuss the relatively large resistance reduction, which appears at the spin flop field of $\simeq 30$~T.

The blue curve in Fig.\,1 shows $R/R_{0}$ of the same Mn$_2$Au(001) thin film in a second field pulse along the same [110]-direction 4~h later, but this time probed with contacts along the perpendicular [1$\bar{1}$0]-direction: $R(B)$ shows a similar hysteresis as described above for the longitudinal resistance measurements, i.\,e.\,a reduction of the resistance at a magnetic field corresponding to the spin-flop field by a very similar value as discussed above for the probe current $\parallel$~[110].

For further investigation of the field pulse related resistance changes, the field was next applied along the hard [010]- and [100]-directions of the Mn$_2$Au thin films, while probing always with the current direction parallel to [010] (see inset of Fig.\,2). For these hard axis directions of the magnetic field, no spin-flop is possible. Instead, the N{\'e}el vectors of the different AFM domains will rotate smoothly from the easy directions towards an alignment perpendicular to the field pulse direction (hard axis), which they will reach at the spin-flop field. This means, that a continuously growing contribution of the AMR effect to the total resistance change is expected for this field direction, which becomes zero again in zero field with the N{\'e}el vector aligned again along the easy [110] and equivalent directions. Indeed, the dependence of the sample resistance on the current direction as shown in Fig.\,2 is much stronger for the field along the hard [100] direction than for ${\vec B} \parallel [110]$. Although the directional dependence of the OMR, which also contributes, is unknown, this provides evidence for a large AMR in the order of some percent associated with the alignment of the N{\'e}el vector parallel to the hard [100] direction in agreement with our previous calculations \cite{Bod18}.  
Additionally, also for the hard axis field direction, a hysteretic resistance reduction is observed. It amounts to about $1.75$~\%, which is about twice as large as for the easy field direction.

\begin{figure}
\includegraphics[width=\columnwidth]{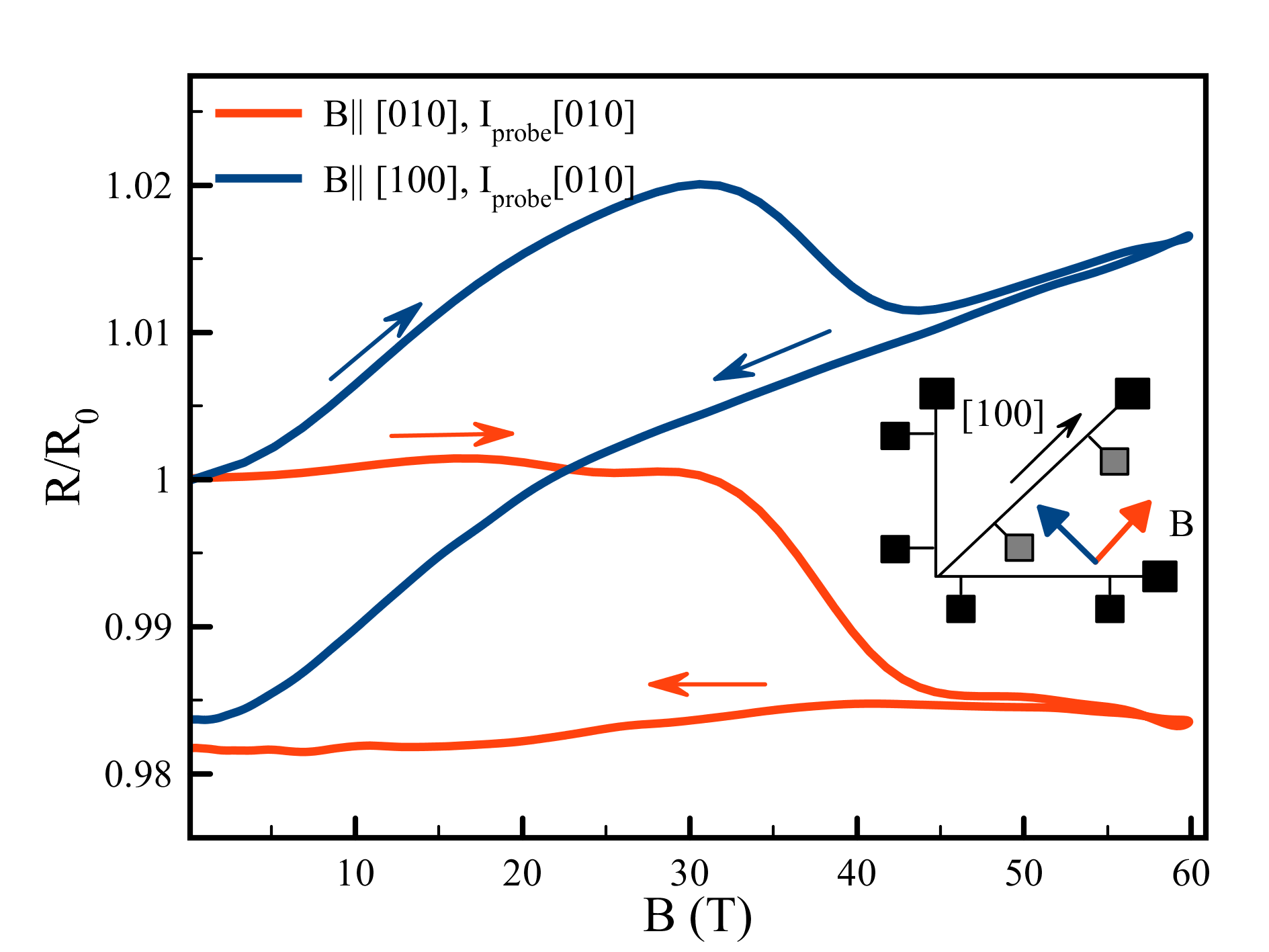}
\caption{\label{Fig2} 
Field dependent resistances of a Mn$_2$Au(001) epitaxial thin film measured during the exposure to a magnetic field pulses along [010] and [100] with a duration of 150~ms, normalized to the sample resistances directly before the application of the respective field pulse. Note that between both field pulses the waiting time was about 4~h.}
\end{figure}
To understand the origin of these changes of the sample resistance, we next probe their temporal stability. These hysteretic resistance reductions are not persistent, i.\,e.\,they relax towards values relatively close to the original resistance of the sample before application of a field pulse:
As shown in Fig.\,3, the resistance reduction generated by the magnetic field exceeding the spin-flop field relaxes for pulses along the hard [010] as well as along the easy [110] direction of the Mn$_2$Au(001) thin films on a time scale of 10~s. 
\begin{figure}
\includegraphics[width=\columnwidth]{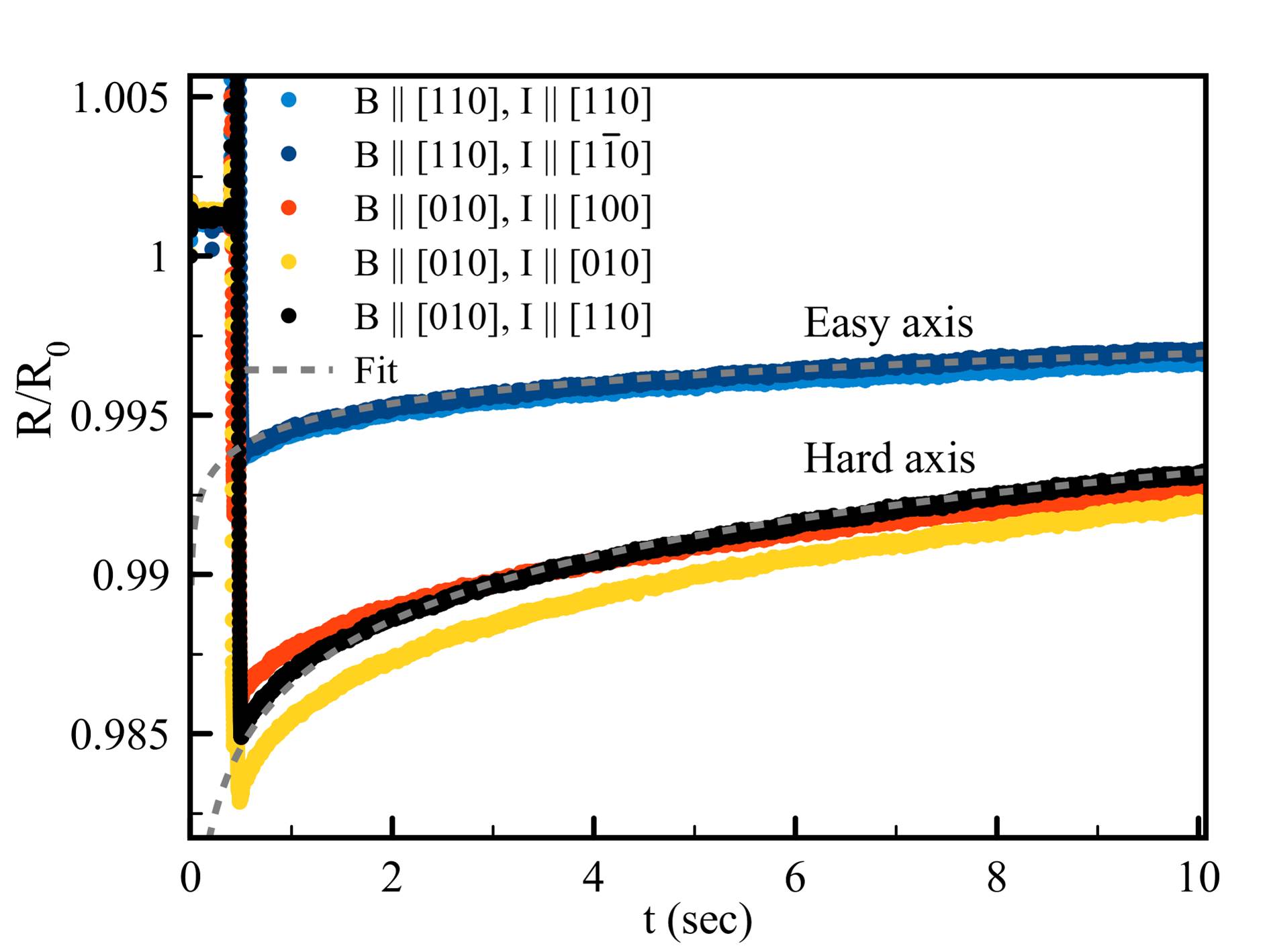}
\caption{\label{Fig3} 
Time dependence of the normalized resistances of a Mn$_2$Au(001) epitaxial thin film probed along different current directions after the exposure to magnetic field pulses along the hard as well as along the easy in-plane axis. The resistance relaxes with a logarithmic time dependence as shown by the fits (dashed curves).}
\end{figure}
Its decay can be fitted by a logarithmic time dependence, which is typical for the relaxation of the magnetization of ferromagnets of different types \cite{Str49,Liu08}. Theoretically, this type of relaxation behavior was first associated with a flat-topped distribution of energy barriers \cite{Ma80}, while later it was shown, that it is rather universal for any distribution of energy barriers \cite{Gon94}.

However, when checking the relaxation of the different signals, we find that not all resistance modifications associated with the field pulse are transient, as we discuss next:

After applying a field pulse along one of the easy axis directions to a sample in the as-grown state, which consists of AFM domains with an equal distribution of domains with the N{\'e}el vector aligned along all four equivalent easy directions (see images in ref.\,\cite{Sap18}), the resistance does not relax completely to its original value. In Fig.\,4, the time dependent sample resistivity is shown for the first 20~min after application of a 60~T field pulse to an as-grown sample, measured by a standard DC-measurement. Please note that with this measurement technique no meaningful data is obtained for the first $\simeq 15$~s after the field pulse due to overloading of the DC nanovoltmeter by the pulse induced induction. A persistent probe current direction ${\bf J}$ dependent increase and decrease of the sample resistance is observed (see inset of Fig.\,4). 
\begin{figure}
\includegraphics[width=\columnwidth]{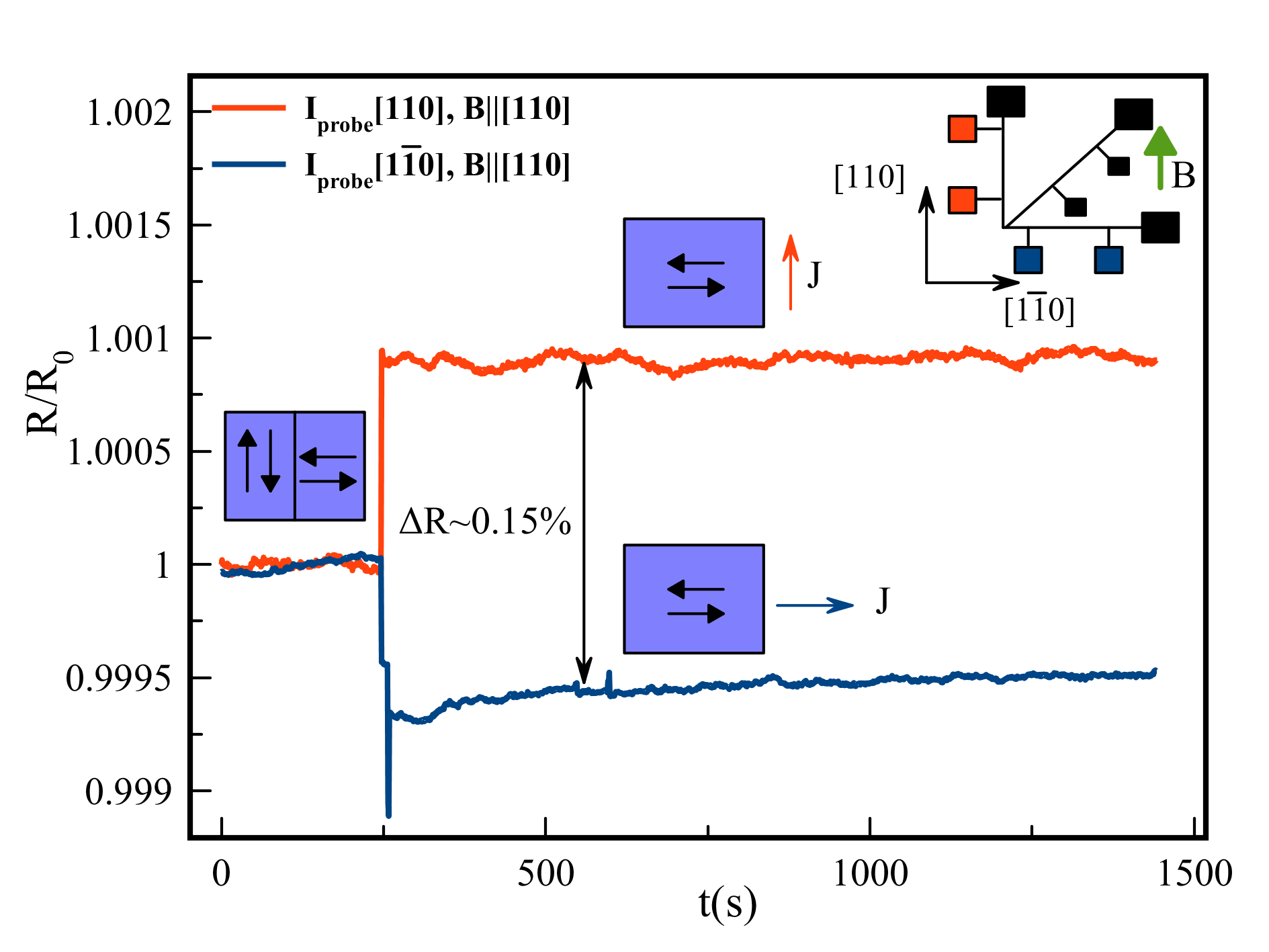}
\caption{\label{Fig5} 
Time dependent resistance for different probe current directions (see inset) of a Mn$_2$Au(001) thin film in the as-grown state and after the application of the first 60~T field pulse at $t\simeq250$~s along the direction indicated in the inset. The standard dc-measurement is unable to produce meaningful data for the first $\simeq 15$~s after the field pulse due to overloading of the nanovoltmeter by the pulse induced induction. The blue boxes indicate the alignment of the N{\'e}el vector.}
\end{figure}
Correspondingly, if the resistance measurement shown in Fig.\,2 is performed for the first time with a sample in the as-grown state and than 4~h later repeated with exactly the same configuration, the initial resistance is changed by exactly the value of the persistent resistance modification discussed here. This long term persistent resistance change of $\Delta R \simeq -0.15$~\% is positive for probe current parallel and negative for probe current perpendicular to the field pulse direction, which corresponds to an AMR effect. An AMR of similar size but with a different sign was recently reported for CuMnAs \cite{Wan20}.

\section{IV. Discussion and Conclusions}
Thus we conclude that in Mn$_2$Au there is a negative AMR of $\simeq -0.15$~\% at 4~K associated directly with the spin-flop driven N{\'e}el vector reorientation. This experimentally observed AMR is in good agreement with our previously published  calculations \cite{Bod18}: we considered the  AMR of Mn$_2$Au(001) theoretically for alignments of the N{\'e}el vector parallel to the [110] as well as parallel to the [100] direction. In the first case, we obtained $AMR_{[110]}\simeq -0.4$~\%, in the second case $AMR_{[100]}\simeq 6$~\%.  As we have meanwhile established, the [110]-direction represents an easy axis within the easy (001)-plane \cite{Sap18}, thus the calculation is in good agreement with the experimental results shown in Fig.\,4. Furthermore, as already discussed above, also the previously calculated $AMR_{[110]}$ in the order of several percent is consistent with the experimentally observed anisotropy of the resistance with the N{\'e}el vector forced by the field pulse to align along the hard axis (Fig.\,2).

However, AMR cannot account for the field pulse generated hysteretic transient resistance reductions, which appear for all field directions at the spin-flop field. This effect can be explained based on DWMR. Independent from the physical mechanism of the DWMR, the large magnetic field during the pulse makes any alignment of the N{\'e}el vector parallel to the field energetically unfavorable. Thus it is likely that during a field pulse with a magnitude above the spin-flop field all domain walls are removed. However, we know from the XMLD-PEEM investigation of our samples performed weeks after the field pulses, that even in samples with a persistent spin-flop induced reorientation of the N{\'e}el vector domain walls exist. We also know that in all configurations the field pulses reduce the resistance of the Mn$_2$Au thin films at the spin-flop field followed by an increase of the resistance after the pulse. Thus we can consistently assume that the domain walls produce a significant DWMR, which is identified by removing the domain walls during the field pulse. After the pulse, on the time scale of 10~s (at 4~K) given by the measured relaxation time of the resistance, the domain walls reform. As the density of the domain walls is very similar in the as grown and in the field manipulated samples \cite{Sap18}, the contribution of the DWMR to the sample resistance is before and after the field pulse is very similar as well. A microscopy based verification of this model is highly desired, but unfortunately not possible as XMLD-PEEM imaging of the domain of Mn$_2$Au cannot be realized on the time scale of 1~s after a 50~T field pulse. Furthermore, we can only speculate about the physical origin of the relatively large DWMR, which is implied by our analysis: We propose, that it is related to the specific anisotropies in the band structure of Mn$_2$Au, from which also the large calculated $AMR_{[100]}\simeq 6$~\% \cite{Bod18} originates. In a domain wall, the N{\'e}el vector is pointing along the hard [100] direction as well and confinement effects could further increase its magnitude. 

Finally, we relate the magnetoresistance effects and magnetic anisotropies discussed here to our previous observation of reversible current pulse induced resistance changes of up to 6~\% \cite{Bod18}: Those large effects where obtained for current pulses along the, as we now know magnetically hard, [100] direction and they were much large as the here experimentally determined persistent $AMR_{[110]}\simeq -0.15$~\%. Thus we can now exclude the AMR of AFM domains with the N{\'e}el vector aligned by current pulses as the origin of these large resistance modifications. However, here we have provided evidence for a potentially large DWMR in Mn$_2$Au, which could well explain the observed resistance changes in the current pulse experiments. Clearly, this requires further investigations combining the microscopic observation of current induced AFM domain manipulation with in-situ resistance measurements. 

\section{V. Summary}
In summary, we have identified the magnitude of the AMR $\simeq -0.15$~\% in Mn$_2$Au associated with the persistent alignment of the N{\'e}el vector along the magnetic easy  [110] axis. Forcing the N{\'e}el vector to align parallel to the hard [110] axis, we provided experimental evidence for an AMR of several percent for this orientation of the staggered magnetization. These experimental values are consistent with our previous calculations of the AMR \cite{Bod18}. Furthermore, we observed field pulse induced transient resistance reductions of $\simeq 1$~\%, which are consistent with the assumption of removal and subsequent reformation of AFM domain walls. This model implies a relatively large domain wall magnetoresistance in Mn$_2$Au. Such large domain wall resistances would provide opportunities in antiferromagnetic spintronics such as race track memory \cite{Yan15} or multi-level switching \cite{Ole17} and could explain the large resistance changes generated by current pulses applied to our Mn$_2$Au(001) thin films \cite{Bod18}. 
\vspace{\baselineskip}\\

\section{ACKNOWLEDGMENTS}
Funded by the Deutsche Forschungsgemeinschaft (DFG, German Research Foundation) – TRR 173 – 268565370 (projects A01, A05 and B12). We acknowledge the support of the HLD at HZDR, member of the European Magnetic Field Laboratory (EMFL). 
\vspace{\baselineskip}\\

\end{document}